\documentclass[aps, reprint, nofootinbib]{revtex4-1}
\usepackage{amsmath, latexsym, amssymb, hyperref, graphicx, color, slashed}
\definecolor{nicered}{rgb}{0.7,0.1,0.1}
\definecolor{nicegreen}{rgb}{0.1,0.5,0.1}
\hypersetup{colorlinks, citecolor=nicegreen,linkcolor=nicered}

\begin{document}
 
\title{Is Left-Right Symmetry the Key?$^{\S}$}

\author{Goran Senjanovi\'c  }

\affiliation{GSSI, L'Aquila, Italy\\
ICTP, Trieste, Italy\\
 goran@ictp.it}
 
\begin{abstract}

   In collaboration with Jogesh Pati, Abdus Salam challenged the chiral gauge nature of the Standard Model by paving the road towards the Left-Right symmetric electro-weak theory. I describe here the logical and historical construction of this theory, by emphasising the pioneering and key role it played for neutrino mass. I show that it is a self-contained and predictive model with the Higgs origin of Majorana neutrino mass, in complete analogy with the SM situation regarding charged fermions. 
  
\end{abstract}


\maketitle

{
 \renewcommand{\thefootnote}%
   {\fnsymbol{footnote}}
 \footnotetext[4]{To appear in the proceeddings of the Abdus Salam Memorial meeting, Singapore, January 2016}
}

\section{Prologue} \label{Prologue}

     It was during my undergraduate days in Beograd, in the early seventies, that I first heard of Abdus Salam as the founder and director of the International Centre for Theoretical Physics in Trieste. For most Yugoslavs Trieste at that time was mainly a shopping mecca of the West - but for a few of us it was becoming a mecca of physics. I found it moving that in a small town at the border with the Balkans Salam would build a place that was supposed to bridge science and scientists from all worlds, especially developing ones. ICTP had no PhD program though and so, following the footsteps of my older brother, I went to do my PhD at the City College of New York. 
          
     CCNY was an exciting place in the seventies. Bob Marshak, one of the fathers of the V-A theory, had  become president and started to build strong physics research program from scratch.      
     And those times in particle physics were great. A whole world of light and excitement opened through the advancement of asymptotic freedom and renormalisability of the spontaneously broken gauge theories. There were fundamental theories and new techniques to learn and one could be a part of history in making. 

I fell in love with the Standard electro-weak model but hated its left-right asymmetry which was supposed to remain for all seasons. It all made sense when Salam, in collaboration with Jogesh Pati, suggested that the world ought to be left-right symmetric in the context of their quark-lepton unification. Pati and Rabi Mohapatra worked out then the minimal Left-Right model, however with Left-Right symmetry broken explicitly, albeit softly. This was troublesome and needed a cure in order that the theory be widely accepted. And so soon after, Mohapatra and I managed to show that parity could be broken spontaneously, thus completing 
 the theory that remains to this day a serious candidate for the physics beyond the Standard Model. 
             
         There was a dark cloud on the horizon though: the theory was predicting massive neutrino against the common wisdom of the day. The trouble was the smallness of neutrino mass, which was hard to understand given that neutrino was predicted to be a Dirac particle, just like the electron. It took some time, but with the advent of the seesaw mechanism and the emergence of the solar neutrino puzzle, neutrino mass became a blessing. 
   The essential point was the necessary existence of the right-handed neutrino which in the seesaw picture becomes a heavy neutral Majorana lepton leading to a light Majorana neutrino. The smoking gun signature is lepton number violation, manifested through neutrinoless double beta decay at low energies, and the production of same sign charged di-leptons at hadronic colliders, as Wai-Yee Keung and I showed in the early eighties. 
         
     \paragraph*{Disclaimer.} This short review emphasises mostly qualitative features and historic developments, and needless to say strong personal bias. I discuss only the what I consider most essential and generic aspects of the minimal Left-Right symmetric theory and the references are rather incomplete. The physics discussed here has been recently reviewed in more depth in~\cite{Senjanovic:2016pza} where one can also find a more complete set of references. 

 \section{General view} \label{General view}
 
    It is clear from the Prologue that the LR symmetric theory~\cite{Pati:1974yy,Mohapatra:1974gc,
    Senjanovic:1975rk,Senjanovic:1978ev} is rather old, so why talk about it today? The answer is twofold. First, finally the LHC has a potential of observing it which makes it more timely than ever. Second, there have been two fundamental developments in recent years that make the theory self-contained and predictive: (i) the testable Higgs origin of neutrino mass~\cite{Nemevsek:2012iq}, in complete analogy with the Standard Model case of charged fermions (ii) the analytic expression for the right-handed quark mixing matrix~\cite{Senjanovic:2014pva}, a challenge that lasted some forty years. Moreover, there has been a furry of activity devoted to the LHC potential and the low energy processes such as neutrinoless double beta decay, lepton flavor violation and such. 
   
       I give the main results at the outset in order to ease the pain for the casual reader and to motivate her to keep reading on.
    
     (i) In the SM the Higgs boson decay rates are completely determined by the masses of particles in question. This is crux of the Higgs mechanism, completed by 
     Weinberg~\cite{Weinberg:1967tq} and GIM~\cite{Glashow:1970gm}.
        In particular, the one-to-one
  correspondence between masses and Yukawa couplings of charged fermions allows one to predict the Higgs boson decays into fermion anti-fermion pairs 
  \begin{equation}\label{eqHiggsdecay}
\Gamma (h \to f \bar f) \propto m_h \frac {m_f^2}{M_W^2}.
 \end{equation}
This is what it means to understand the origin of particle masses. One can worry why the masses are what they are, but this question, if it is ever to be answered, comes after one establishes their Higgs-Weinberg origin.

   It is in this sense that the LR symmetric model is the theory of neutrino mass, as will be discussed in section \ref{LRSM is a theory of neutrino mass}. In direct analogy  with \eqref{eqHiggsdecay} one can predict~\cite{Nemevsek:2012iq} the Higgs decay into light and heavy neutrinos, or better, the decay of heavy right-handed neutrino $N$ (when it is heavier than the Higgs) into the Higgs and light neutrino. As an illustration, I give here the relevant expression~\cite{Tello:2012qda} for a simplified case described in the section \ref{LRSM is a theory of neutrino mass}
  \begin{equation}\label{eqN2hdecay}
\Gamma (N_i \to h \nu_j) \propto \delta_{ij} \,\,m_{\nu_i} \frac{m_{N_i}^2}{M_W^2}.
 \end{equation}
%
This would be hard to observe, needles to say; however, there is an experimentally more accessible decay channel of right-handed neutrino N into the W boson and charged lepton~\cite{Nemevsek:2012iq,Tello:2012qda} 
  \begin{equation}\label{eqNdecay}
\Gamma (N_i \to W \ell_j) \propto V_{i j}^2 m_{\nu_i} \frac{m_{N_i}^2}{M_W^2}.
 \end{equation}
where $V$ is the PMNS leptonic mixing matrix.
    
    In the general case the above expressions look somewhat more complicated, but all the essential features are caught here. One has a complete analogy with the Standard Model situation regarding the charged fermions, only now one has to know the (Majorana) masses and mixings of left and right handed neutrinos separately. More about it in the section \ref{LRSM is a theory of neutrino mass}.       
   
   (ii) The right-handed quark mixing matrix $V_R$ has a simple approximate  form~\cite{Senjanovic:2014pva} as a function of the usual left-handed CKM matrix $V_L$
\begin{equation}
\label{eq:VR}
(V_R)_{ij} \simeq (V_L)_{ij} - i  \epsilon  \frac{(V_L)_{ik} ( V_L^{\dagger}m_uV_L)_{kj} }{m_{d_k}+m_{d_j}}  
+O(\epsilon^2) 
\end{equation}
where $\epsilon $ is a small unknown expansion parameter. It can be shown that the left and right mixing angles are almost the same, and right-handed phases depend only on $V_L$ and  
$\epsilon $. A determined reader should go to the section \ref{LRSM is a theory of RH quark mixing angles}
 for more details and for some immediate consequences of \eqref{eq:VR} regarding the right-handed mixing angles and phases.

   The rest of this short review is organised as follows. I first discuss the salient features of the theory in the next section, and then try to give a historical development that took one to the seesaw based version of the model. Thus, in the section \ref{Classic era} I go through the original version of theory that had Dirac neutrinos and struggled explaining why their masses were so small. The section \ref{LR theory: modern era} is devoted to the modern version of the theory based on the seesaw mechanism with naturally light Majorana neutrinos.
   Next, I go over the issues (i) and (ii) above in the sections \ref{LRSM is a theory of neutrino mass} and \ref{LRSM is a theory of RH quark mixing angles}, respectively, before offering an outlook for the future. I end with an epilogue, in order to make the presentation not only LR but also top-bottom symmetric.

    \section{Generic features} \label{LR theory: generic features}

       The minimal LR symmetric theory is based on the  
$ SU(2)_L \times SU(2)_R \times U(1)_{B-L}$ gauge group, augmented by 
the symmetry between the left and right sectors~\cite{Pati:1974yy,Mohapatra:1974gc,
    Senjanovic:1975rk,Senjanovic:1978ev}. 

    Quarks and
leptons are then completely LR symmetric
\begin{equation}
q_{L,R }= \left( \begin{array}{c} u \\ d \end{array}\right)_{\!\! \!\!L,R}\,,\qquad
\ell_{L,R} = \left( \begin{array}{c} \nu \\ e \end{array}\right)_{\!\!\!\! L,R}.
\label{ds21}
\end{equation}

   Clearly, the LR symmetry says that if there is a LH neutrino, there must be the RH one too and neutrino cannot remain massless. A desire to cure the left-right asymmetry of weak interactions lead automatically to neutrino mass.
  
The formula for the electromagnetic charge becomes~\cite{Mohapatra:1980qe}
\begin{equation}
Q_{em} = I_{3 L} + I_{3 R} + {B - L \over 2}\,.
\label{ds22}
\end{equation}
which trades the hard to recall hyper-charge of the SM for $B-L$, the physical anomaly-free global symmetry of the SM, now gauged. Both LR symmetry and the gauged $B-L$ require the presence of RH neutrinos. 

\paragraph*{LR symmetries.}  It is easy to verify that the only realistic discrete LR
symmetries, preserving the kinetic terms, are
 $\mathcal{P}$ and $\mathcal{C}$, the generalised parity and charge-conjugation respectively, supplemented by the
exchange of the left and right $SU(2)$ gauge groups (for a recent discussion and references, 
see~\cite{Dekens:2014ina}).

\paragraph*{Higgs sector.} The analog of the SM Higgs doublet is now a bi-doublet~\cite{Mohapatra:1974gc,
    Senjanovic:1975rk,Senjanovic:1978ev}
\begin{equation}\label{eq:bidoublet}
\Phi = \left[\begin{array}{cc}\phi_1^{0}&\,\,\,\, \phi_2^+\\ \phi_1^-&-\phi_2^{0*}\end{array}\right]
\end{equation}
in order to provide masses for charged fermions. This amounts to two $SU(2)_L$ doublets, but one of them ends up being very heavy and effectively decouples from low energies~\cite{Senjanovic:1978ev}. In analogy with their charged partners neutrinos get Dirac mass.

\section{Classic era} \label{Classic era}

 So far so good. But what fields should be used for the large scale of symmetry breaking? 

In the original version~\cite{Mohapatra:1974gc,
    Senjanovic:1975rk,Senjanovic:1978ev} one opted for  $B-L= 1$ LH and RH doublets, i.e. the doublets under $SU(2)_L$ and $SU(2)_R$ groups respectively. It seemed a logical choice, a LR extension of the SM Higgs doublet. Looking back, it is hard to understand why for some years no alternative was studied, since there was nothing special about this choice. After all, we had already used the $SU(2)_L$ doublets, in a form of a bi-doublet, in order to give masses to charged fermions. The large scale Higgs sector determines the ratio of new gauge boson masses, $W_R$ and $Z_R$ but why in the world should it mimic the SM situation with $W$ and $Z$? It is both instructive and amusing how one gets sidetracked and confused at the beginnings.

    In any case, one sat down to show that parity could be broken spontaneously~\cite{Senjanovic:1975rk,Senjanovic:1978ev}. The symmetry of the potential tells you immediately that there are only two possibilities:
    (i) same vevs for LH and RH doublets, unacceptable; (ii) one of the two vevs vanishing, as required by experiment. 
     It was easy to show that there was the stable minimum with only the RH doublet vev, and the task of breaking the theory down to the SM was achieved~\cite{Senjanovic:1975rk,Senjanovic:1978ev}. 
     
     The main prediction of the model was a massive neutrino, a Dirac fermion just like the electron. So why in the world was it so light?

\section{Modern era} \label{LR theory: modern era}

  So, the theory was prophetic in predicting neutrino mass so early, years before experiment, but it seemed to fail to account for its smallness~\cite{Branco:1978bz}. 
   It turned out that the problem was not LR symmetric gauge group, but simply the choice of the heavy Higgs sector. This let to the with a version of the theory based on the seesaw mechanism~\cite{Minkowski:1977sc, Mohapatra:1979ia,seesaw}.
       
    The main point was to chose the right Higgs in order to make RH neutrino a heavy Majorana lepton, so that it could mix with the LH one and give it in turn a tiny mass. All that it required was to substitute doublets by the appropriate triplets. In this way, the theory led naturally to a Majorana neutrino and lepton number violation (LNV). In our paper~\cite{Mohapatra:1979ia}, Mohapatra and I emphasised this, and argued that the resulting LNV process, the neutrinoless double decay~\cite{Racah:1937qq}, could be easily generated by new RH sector, and not by small Majorana neutrino mass. Yet, it is often claimed to this day that this process is a direct probe of neutrino Majorana mass; this is simply wrong. 
     Btw, the idea of new physics being possibly behind the neutrinoless double beta decay dates back all the way to the late fifties~\cite{Feinberg:1959}.    
        
    Some years later Wai-Yee Keung~\cite{Keung:1983uu} and I made a case for an analog high-energy Lepton Number Violating process, the production and the subsequent decay of the RH neutrino. We realised that due to its Majorana nature, the RH neutrino, once produced on-shell would decay equally into a charged lepton and anti-lepton. This allows to test and measure directly its Majorana nature, not just indirectly through low energy effective processes. Also, besides the usual LNV conserving final state, one would have direct LNV in the form of the same sign charged di-leptons and (two) jets. This turns out to be a generic property of any theory that leads to Majorana neutrino and has become over the years the paradigm for LNV at hadronic colliders, and today both CMS and ATLAS are looking into it. 
        
    What follows is a brief, almost telegraphic review of this subject.   A reader that wishes to dig deeper can consult more detailed recent overviews in~\cite{Senjanovic:2010nq,Tello:2012qda} or a classic book on the subject~\cite{Mohapatra:1998rq}, especially regarding the neutrino stuff.
  
 In summary, the modern day version of the theory is based on the seesaw mechanism. The Higgs sector consists of the following multiplets~\cite{Mohapatra:1979ia,Minkowski:1977sc}: the bi-doublet $\Phi$ of \eqref{eq:bidoublet} and the
$SU(2)_{L,R}$ triplets $\Delta_{L,R}$ 
\begin{equation}
\Delta_{L, R} = \left[ \begin{array}{cc} \Delta^+ /\sqrt{2}& \Delta^{++} \\
\Delta^0 & -\Delta^{+}/\sqrt{2} \end{array} \right]_{\!\!L,R}
\label{ds32}
\end{equation}
The first stage of symmetry breaking down to the SM symmetry
takes the following form~\cite{Senjanovic:1975rk,Senjanovic:1978ev, Mohapatra:1979ia}
\begin{equation}
  \langle \Delta_L^0\rangle = 0, \qquad \langle \Delta_R^0\rangle = v_R
\end{equation}
with $v_R$ giving masses to the heavy charged and neutral gauge bosons $W_R, Z_R$, right-handed neutrinos 
and all the scalars except for the usual Higgs doublet (the light doublet in the bi-doublet $\Phi$).
Next, the neutral components of $\Phi$ develop vevs and break the SM symmetry down to $U(1)_{em}$
\begin{equation}\label{eq:Phivev}
\langle\Phi\rangle=v\, \text{diag} (\cos\beta,-\sin\beta e^{-ia})
\end{equation}
where $v$ is real and positive and $ \beta<\pi/4$, $0 < a < 2 \pi$. 

In turn, $\Delta_L$ develops a tiny induced vev 
$\langle\Delta_L\rangle\propto v^2/v_R$~\cite{Mohapatra:1980yp} which contributes directly to neutrino mass. Its smallness is naturally controlled by a small quartic coupling, sensitive only to the seesaw contribution~\cite{Mohapatra:1980yp}. 

I should stress that there is confusion to this day regarding the issue of naturalness of small $v_L$, and it is even argued that parity ought to be broken at the high scale (with a gauge singlet) in order to make $\Delta_L$ heavy enough, and effectively decouple it from the physics of the LR theory. However, large scales only add a hierarchy problem and thus make things worse. A small, protected coupling is definitely more natural than a large ratio of mass scales. Moreover, breaking parity through a singlet vev is physically equivalent to the soft breaking, and is a step backward towards the original formulation when it was claimed that parity had to be broken softly. 

The soft breaking (or the large scale spontaneous breaking) alleviates the infamous domain wall problem~\cite{Zeldovich:1974uw} of spontaneously broken discrete symmetries, but this may not be such a problem after all. It turns out that even tiny symmetry breaking gravitational effects suppressed by the Planck scale suffice to destabilise the domain walls~\cite{Rai:1992xw}. Also, symmetry non-restoration at high temperature~\cite{Weinberg:1974hy} may offer an alternative way out~\cite{Dvali:1995cc}. 

This said, the breaking of LR symmetry, but only when it is generalised charge conjugation $\mathcal{C}$, can happen naturally in grand unified theories~\cite{Chang:1983fu}. However, in minimal predictive grand unified theories such as $SO(10)$, $M_R$ ends up being enormous, on the order of $10^{10} \,\text{GeV}$~\cite{delAguila:1980qag}.

Before we move on, a comment is called for. By the early eighties the LR theory was fully developed, and yet most of us stopped working on it until the LHC came along. The reason is that it became clear already in 1981 that the LR scale had to be large, on the order of few $\text{TeV}$, from the $K_L-K_S$ mass difference~\cite{Beall:1981ze}. Huge, but not out of the LHC reach of about $5-6\,\text{ TeV}$~\cite{Ferrari:2000sp, Gninenko:2006br}. Actually, LHC has already caught up with the theoretical limit for a large portion of the parameter space of the RH neutrino masses~\cite{Khachatryan:2014dka}.

\section{LRSM is a theory of neutrino mass} \label{LRSM is a theory of neutrino mass}

    Let us see more carefully what happens with neutrino mass in this theory, and how we could probe directly its origin. 
 
\begin{figure}
\centering
\includegraphics[scale=0.9]{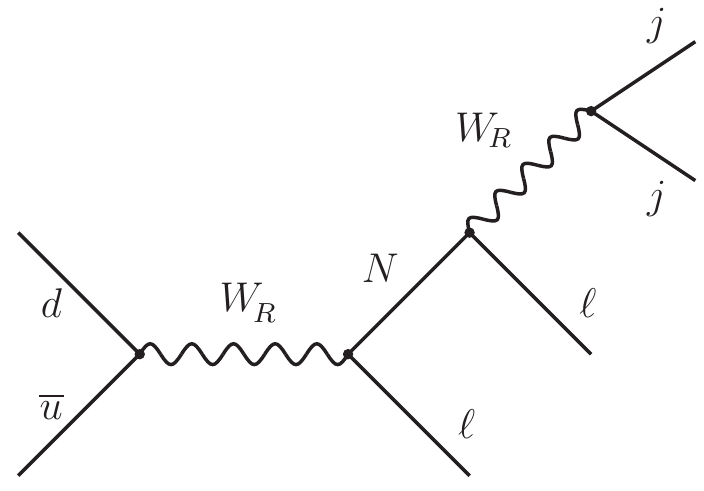}
\caption{The KS production process of lepton number violating same sign di-leptons through the production and subsequent decay of $N$.  }\label{KS}
\end{figure}

    The simple thing to realise is that now we need to measure both LH and RH neutrino masses and mixings. We are slowly but surely doing the job for the light neutrinos and it is only a matter of time to complete it. In the case of RH neutrinos, we need to produce them at colliders, and LHC is the custom-fit machine for this, with spectacular manifestation of the LNV in the form of same sign charged di-lepton pairs accompanied by two jets~\cite{Keung:1983uu},
shown in the Fig.~\ref{KS}.

This process allows for the possibility of establishing directly the Majorana nature of $N$ since then both same and opposite sign charged leptons decay products occur 
 with the same probability. It should be stressed that this has become the paradigm for LNV at the hadronic colliders, and it occurs in basically any theory that leads to Majorana neutrinos. Moreover, there is a deep connection between 
lepton number violation at LHC and in neutrinoless double decay~\cite{Tello:2010am}.

     In the LR model the dominant LNV effect is through the on-shell production of $W_R$; it could also occur through the small $\nu-N$ mixing and the usual $W$ exchange, but that requires huge $M_D$~\cite{Pilaftsis:1991ug}. In this manner, the smallness of neutrino mass would be a complete accident, nothing to do with the seesaw.

In the limit of small $v_L$, chosen only for the sake of illustration, the Majorana neutrino mass matrix is given by the usual seesaw expression
\begin{equation} \label{eqSeesaw}
  M_\nu = - M_D^T \frac{1}{M_N} M_D,
\end{equation}
where $M_D$ is the neutrino Dirac mass matrix, while $M_N\propto M_{W_R}$ is the symmetric Majorana mass matrix right-handed neutrinos. The smallness of neutrino mass is the consequence
 of near maximality of parity violation in beta decay, and in the infinite limit for the $W_R$ mass
 one recovers massless neutrinos of the SM.

The case of $C$ as the LR symmetry is rather illustrative, since it implies symmetric Dirac mass matrix $M_D = M_D^T$, which eliminates the arbitrary complex orthogonal matrix $O$ that obscures~\cite{Casas:2001sr} the usual seesaw mechanism of the SM with $N$. This provides the fundamental difference between the naive seesaw and the LR symmetric theory, since in LR  Dirac mass matrix $M_D$ can be obtained~\cite{Nemevsek:2012iq} directly from~\eqref{eqSeesaw} 
\begin{align} 
\label{eqMD}
&  M_D = i \, M_N \sqrt{M_N^{-1} M_\nu},
\end{align}
and thereby one can determine the mixing between light and heavy neutrinos. 

I cannot overstress the importance of this result. One often invokes discrete symmetries in order to obtain information on Dirac Yukawa couplings, but this is completely unnecessary since the theory itself predicts it, just as the knowledge of charged fermion masses predicts the corresponding Yukawas in the SM. The LR model is a self-contained predictive theory of neutrino mass, as title of this sections says.

The crucial thing is that $N$, besides decaying through virtual $W_R$ as discussed above, decays also into the left-handed charged lepton through $M_D/M_N$ 
\cite{Buchmuller:1991tu,Pilaftsis:1991ug}.
In a physically interesting case when $N$ is heavier than $W_L $, the decay into left-handed leptons proceeds through the on-shell production of $W_L$. For the sake of illustration we choose an example of $V_R=V_L^*$,
which leads to the expression for the $N \to W \ell$ decay given in \eqref{eqNdecay} of the Introduction. 
That is the analog of ~\eqref{eqHiggsdecay} which probes the Higgs origin of charged fermion masses; while in the SM it suffices to know $m_f$, in the case of neutrinos one needs to know both light and heavy masses and mixings.
%

For this reason the $N \to W \ell$ decay is physically more relevant. We saw from \eqref{eqNdecay} that the corresponding decay width is very small, so it could appear hopeless to be observed. The main decay of $N$ proceeds through the $W_R$ channel for the LHC relevant mass scale and it is quite slow too. Thus the branching ratio for $N \to W \ell$ decay is not necessarily negligible; one finds for 
  the ratio of $N$ decays in the $W_L$ and $W_R$ channels~\cite{Nemevsek:2012iq}
\begin{equation}
	\frac{\Gamma_{N \to \ell_L j j}}{\Gamma_{N \to \ell_R j j}} \simeq 10^3 \frac{M_{W_R}^4 m_\nu}{M_{W_L}^2 m_N^3},
\end{equation}
which is maximally about a per-mil for $W_R$ accessible at the LHC.

The situation in the case of $P$ is more subtle and less illuminating. Suffice it to say that the end result is basically the same and once again Dirac Yukawas get determined as the function of light and heavy neutrino mass matrices.

%
%


\section{LRSM is a theory of RH quark mixing angles} \label{LRSM is a theory of RH quark mixing angles}

   In the perfectly LR symmetric world, the LH and RH mixing matrices would be the same, but we live in a badly broken symmetry world. Does this mean that the RH quark mixing matrix $V_R$ is not predicted at all? The answer, surprisingly, is negative: the RH quark mixing angles are predicted by the theory, and when LR symmetry is generalised parity the RH phases get determined too. 
   
     The case of generalised charge conjugation is easier to discuss, albeit less exciting. The Yukawa matrices are symmetric, and thus are also quark mass matrices. In turn, this implies same LH and RH mixing angles, with new arbitrary phases in the $V_R$ matrix. End of story.
     
      The case of parity is more interesting and has a long history. Yukawa matrices are hermitean, but the complex vev of \eqref{eq:Phivev} destroys the hermiticity of the quark mass matrices. However, spontaneous breaking is soft and to some degree keeps the memory of the original symmetric world. One could easily imagine that $V_R$ may be related to $V_L$, but the actual analytical computation was a great challenge for some forty years. Numerical calculations indicated that the LH and RH mixing angles were similar, but the first serious attempt to compute $V_R$ was made only some ten years ago~\cite{Zhang:2007fn}, in a limited portion of parameter space and not as clearly as one would have hoped for direct application. The task was finally completed two years ago,
      when the analytic form valid in the entire parameter space was finally obtained~\cite{Senjanovic:2014pva}
The leading form is given by simple \eqref{eq:VR}, used in Introduction to anticipate the discussion. For a complete form and extensive discussion see~\cite{Senjanovic:2014pva} where the leading terms are derived for the differences between mixing angles
    \begin{align}\label{eq:difference12}
   \theta^{12}_R - \theta^{12}_L &\simeq  - s_a t_{2\beta} \frac {m_t}{m_s} s_{23} s_{13}   s_\delta
\\[3pt]
   \label{eq:difference23}
   \begin{split}
\theta^{23}_R - \theta^{23}_L &\simeq -  s_a t_{2\beta}  \frac{m_t}{m_b}\frac{m_s}{m_b} s_{12} s_{13} s_\delta
\end{split}
\\[3pt]
    \label{eq:difference13}
    \begin{split}
     \theta^{13}_R - \theta^{13}_L &\simeq - s_a t_{2\beta} \frac{m_t}{m_b} \frac{m_s}{m_b} s_{12} s_{23} s_\delta\end{split}
 \end{align}
 and similarly for the KM phases
\begin{align} 
   \label{eq:Delta}
  \delta_R - \delta_L  &\simeq  s_a t_{2\beta} \frac{m_c +m_t s_{23}^2}{m_s}
  \end{align}
 where, for simplicity, we defined $s_{ij}=\sin \theta^L_{ij}$ and $s_\delta= \sin \delta_L$ and a and $\beta$ are defined in \eqref{eq:Phivev}.
   It should be kept in mind that the phase difference $\delta_R - \delta_L$ is always accompanied with the factor $\sin\theta^{13}_L$.

 The LH and RH mixing angles are almost exactly the same. A surprising result in view of the fact that parity is maximally broken at low energies? Well, partially. The breaking of parity is spontaneous, so the memory of it remains to some degree. Still this by itself is not sufficient and the above result is to a large degree due to the smallness of the LH mixing angles. We know that nature conspires to make the SM work so well, CKM mixings must be small, and the same works nicely for the LR model, preserving the LR symmetry between the mixings.
 
  Another important result: the new RH phases depend~\cite{Senjanovic:2014pva} on a single parameter $s_a t_{2\beta}$ which measures the departure from the hermiticity of quark mass matrices. 
  
      \section{Summary and outlook} \label{Summary and outlook} 
      
        Today, some forty years after its construction, the LR symmetric theory is as sound as ever and finally with a potential of being experimentally probed. The theoretical limits from the early eighties told us to wait for the LHC in order to start testing it and finally this possibility has arrived. The smoking gun signature of the theory is the production of the RH charged gauge boson and its subsequent decay into RH neutrinos and charged leptons. The Majorana nature of RH neutrinos then predicts equal amount of same and opposite sign di-leptons~\cite{Keung:1983uu}. On top of direct lepton number violation at hadronic colliders such as the LHC one has a unique opportunity to verify the Majorana property of RH neutrinos. In turns out that, through the predicted neutrino Dirac mass matrix~\cite{Nemevsek:2012iq}, one has a possibility of probing the seesaw mechanism in the context of the LR theory, as opposed to the situation in the SM augmented with the seesaw.
        
            This may sound nice but the reader must be asking herself as to why in the world should the LR scale be accessible to present day accelerators? After all, it be easily close to the GUT scale and yet give observable neutrino mass. I have no way of defending the tempting desire to see the restoration of parity in a foreseeable future. There is however a possible phenomenological motivation to pursue this: the neutrinoless double beta decay. If it were to be observed and neutrino mass not sufficient to account for it, new physics would be a must. It is easy to see that this would require the scale of new physics not to be much above $10\, \text{TeV}$ or so. With the LHC reach close to that, it is imperative to be ready for such a possibility and study the consequences described here. I should add that there are a number of other phenomenological possibilities, both in gauge and scalar sectors, including lepton number violation leaking into the SM Higgs decays~\cite{Gunion:1986im}.
            
              Left-Right symmetry however may choose to reveal itself at the future collider (for a recent discussion see~\cite{Dev:2016dja}). The heavy Higgs doublet whose couplings violate flavor must lie above $20\, \text{TeV}$ or so~\cite{Zhang:2007da, Blanke:2011ry} which implies that the corresponding coupling is not far from its perturbativity limit~\cite{Maiezza:2016bzp}. Once the $W_R$ mass is raised above $10\, \text{TeV}$, the situation improves and around $20\, \text{TeV}$, the theory is perfectly perturbative with the large strong coupling cut-off scale. And in the case of LR symmetry being parity, this would ensure that the strong CP parameter is naturally small~\cite{Maiezza:2014ala}.
             
             The new generation of hadronic colliders would have a serious chance of observing LR symmetry if we were to see the neutrinoless double beta decay and know that it is due to new physics (for a roadmap, 
             see~\cite{Nemevsek:2011hz}). One could probe both $W_R$ and $N$ masses~\cite{Keung:1983uu}, the RH leptonic mixing angles~\cite{Das:2012ii,Vasquez:2014mxa}, measure the chirality of $N$ couplings, establish their RH nature~\cite{Ferrari:2000sp,Han:2012vk} and study in detail their Majorana character~\cite{Gluza:2016qqv}. 
 We could truly probe the nature and the origin of neutrino mass~\cite{Nemevsek:2012iq} just as we are doing now for charged fermions.

      \section{Epilogue}  \label{Epilogue}  
      I started this short review with a personal account of the seventies when the budding LR theory was emerging as one of the leading candidates for BSM physics. 
                   Years passed. In the eighties I often lectured at the Trieste summer schools where I got to know Salam well and fell in love with the ICTP and its mission. Some years later I went there to build the phenomenology group, then non-existent. Today, more than a quarter of a century later I look back with pride and joy at having had a role in history in the making. Having had a great number of collaborators and Diploma and PhD students from developing countries, I wish to express my happiness and gratitude to have been involved so deeply with this tremendous project. 
 
 From where I stand today, Salam died quite young, only four years older than I am now. Losing Salam was very hard on our High Energy group and all of ICTP. For this reason, I am happy to add my voice to the tribute of a great physicist who left us all too soon.

  \section{Acknowledgments}
  
     I wish to thank the organisers of the Abdus Salam Memorial meeting for an offer to contribute to the proceedings in spite of having been unable to be present in Singapore. I am grateful to Alejandra Melfo and Vladimir Tello for discussions and help in improving the physics and style presentation of this manuscript.


\begin{thebibliography}{99}


\bibitem{Senjanovic:2016pza} 
  G.~Senjanovi\'c and V.~Tello,
  ``Origin of Neutrino Mass,''
  PoS PLANCK {\bf 2015}, 141 (2016).
  

%
\bibitem{Pati:1974yy} 
  J.~C.~Pati and A.~Salam,
  ``Lepton Number as the Fourth Color,''
  Phys.\ Rev.\ D {\bf 10}, 275 (1974)
  Erratum: [Phys.\ Rev.\ D {\bf 11}, 703 (1975)].
  doi:10.1103/PhysRevD.10.275, 10.1103/PhysRevD.11.703.2

\bibitem{Mohapatra:1974gc} 
  R.~N.~Mohapatra and J.~C.~Pati,
  ``A Natural Left-Right Symmetry,''
  Phys.\ Rev.\ D {\bf 11}, 2558 (1975).
  doi:10.1103/PhysRevD.11.2558
 
\bibitem{Senjanovic:1975rk}
G.~Senjanovi\'c and R.~N.~Mohapatra,
``Exact Left-Right Symmetry And Spontaneous Violation Of Parity,''
Phys.\ Rev.\ D {\bf 12} (1975) 1502.


\bibitem{Senjanovic:1978ev} 
  G.~Senjanovi\'c,
  ``Spontaneous Breakdown of Parity in a Class of Gauge Theories,''
  Nucl.\ Phys.\ B {\bf 153}, 334 (1979).
  doi:10.1016/0550-3213(79)90604-7
  

\bibitem{Nemevsek:2012iq}
  M.~Nemev\v{s}ek, G.~Senjanovi\'c and V.~Tello,
  ``Connecting Dirac and Majorana Neutrino Mass Matrices in the Minimal Left-Right Symmetric Model,''
  Phys.\ Rev.\ Lett.\  {\bf 110} (2013) 15,  151802
  [arXiv:1211.2837 [hep-ph]].

Notice that the title in the published version is different (and much less reader friendly) from the original one in the arXiv (courtesy of PRL). Looking back, a title such as ``Higgs Origin of Majorana Neutrino Mass'' could have probably helped better to make the case.
  
\bibitem{Senjanovic:2014pva} 
  G.~Senjanovi\'c and V.~Tello,
  ``Right Handed Quark Mixing in Left-Right Symmetric Theory,''
  Phys.\ Rev.\ Lett.\  {\bf 114}, no. 7, 071801 (2015)
  [arXiv:1408.3835 [hep-ph]].

  G.~Senjanovi\'c and V.~Tello,
  ``Restoration of Parity and the Right-Handed Analog of the CKM Matrix,''
  arXiv:1502.05704 [hep-ph].
  

  
\bibitem{Weinberg:1967tq} 
  S.~Weinberg,
  ``A Model of Leptons,''
  Phys.\ Rev.\ Lett.\  {\bf 19}, 1264 (1967).
  doi:10.1103/PhysRevLett.19.1264
  
\bibitem{Glashow:1970gm} 
  S.~L.~Glashow, J.~Iliopoulos and L.~Maiani,
  ``Weak Interactions with Lepton-Hadron Symmetry,''
  Phys.\ Rev.\ D {\bf 2}, 1285 (1970).
  doi:10.1103/PhysRevD.2.1285

 
\bibitem{Tello:2012qda}
  V.~Tello, PhD Thesis, SISSA (2012)
  ``Connections between the high and low energy violation of Lepton and Flavor numbers in the minimal left-right symmetric model,''
 
 
\bibitem{Mohapatra:1980qe}
  R.~N.~Mohapatra and R.~E.~Marshak,
  ``Local B-L Symmetry of Electroweak Interactions, Majorana Neutrinos and Neutron Oscillations,''
  Phys.\ Rev.\ Lett.\  {\bf 44} (1980) 1316
   Erratum: [Phys.\ Rev.\ Lett.\  {\bf 44} (1980) 1643].
  doi:10.1103/PhysRevLett.44.1316

 

\bibitem{Dekens:2014ina}
  W.~Dekens and D.~Boer,
  Nucl.\ Phys.\ B {\bf 889} (2014) 727
  doi:10.1016/j.nuclphysb.2014.10.025
  [arXiv:1409.4052 [hep-ph]].
  
%
%
  
\bibitem{Branco:1978bz}
  G.~C.~Branco and G.~Senjanovi\'c,
  ``The Question of Neutrino Mass,''
  Phys.\ Rev.\ D {\bf 18} (1978) 1621.


\bibitem{Minkowski:1977sc}
P.~Minkowski,
``Mu $\to$ E Gamma At A Rate Of One Out Of 1-Billion Muon Decays?,''
Phys.\ Lett.\ B {\bf 67} (1977) 421.


\bibitem{Mohapatra:1979ia}
  R.~N.~Mohapatra and G.~Senjanovi\'c,
  ``Neutrino Mass and Spontaneous Parity Violation,''
  Phys.\ Rev.\ Lett.\  {\bf 44} (1980) 912.
  

\bibitem{seesaw} 
  S.~L.~Glashow,
  ``The Future of Elementary Particle Physics,''
  NATO Sci.\ Ser.\ B {\bf 61}, 687 (1980).

  M.~Gell-Mann, P.~Ramond and R.~Slansky,
  ``Complex Spinors and Unified Theories,''
  Conf.\ Proc.\ C {\bf 790927} (1979) 315
  [arXiv:1306.4669 [hep-th]].

  T.~Yanagida,
  ``Horizontal Symmetry And Masses Of Neutrinos,''
  Conf.\ Proc.\ C {\bf 7902131}, 95 (1979).


 
\bibitem{Racah:1937qq}
  G.~Racah,
  ``On the symmetry of particle and antiparticle,''
  Nuovo Cim.\  {\bf 14}, 322 (1937)

  W.~H.~Furry,
  ``On transition probabilities in double beta-disintegration,''
  Phys.\ Rev.\  {\bf 56}, 1184 (1939).


\bibitem{Feinberg:1959}
	G.~Feinberg, M.~Goldhaber,
	Proc.\ Nat.\ Ac.\ Sci.\ USA {\bf 45} (1959) 1301;
  
  B.~Pontecorvo,
  ``Superweak interactions and double beta decay,''
 Phys.\ Lett.\  {\bf B26 } (1968)  630.


\bibitem{Keung:1983uu}
  W.~Y.~Keung and G.~Senjanovi\'c,
  ``Majorana Neutrinos And The Production Of The Right-Handed Charged Gauge
  Boson,''
  Phys.\ Rev.\ Lett.\  {\bf 50}, 1427 (1983).




\bibitem{Senjanovic:2010nq} 
  G.~Senjanovi\'c,
  ``Seesaw at LHC through Left - Right Symmetry,''
  Int.\ J.\ Mod.\ Phys.\ A {\bf 26}, 1469 (2011)
  [arXiv:1012.4104 [hep-ph]].
  
  G.~Senjanovi\'c,
  ``Neutrino mass: From LHC to grand unification,''
  Riv.\ Nuovo Cim.\  {\bf 034}, 1 (2011).

\bibitem{Mohapatra:1998rq}
  R.~N.~Mohapatra and P.~B.~Pal,
  ``Massive neutrinos in physics and astrophysics. Second edition,''
  World Sci.\ Lect.\ Notes Phys.\  {\bf 60} (1998) 1
   [World Sci.\ Lect.\ Notes Phys.\  {\bf 72} (2004) 1].

\bibitem{Mohapatra:1980yp}
R.~Mohapatra, G.~Senjanovi\'{c},
``Neutrino Masses And Mixings In Gauge Models With Spontaneous Parity
Violation,''
Phys.Rev.{\bf D23} (1981) 165.


\bibitem{Zeldovich:1974uw} 
  Y.~B.~Zeldovich, I.~Y.~Kobzarev and L.~B.~Okun,
  ``Cosmological Consequences of the Spontaneous Breakdown of Discrete Symmetry,''
  Zh.\ Eksp.\ Teor.\ Fiz.\  {\bf 67}, 3 (1974)
  [Sov.\ Phys.\ JETP {\bf 40}, 1 (1974)].
  
  
\bibitem{Rai:1992xw}
  B.~Rai, G.~Senjanovi\'c,
  ``Gravity and domain wall problem,''
  Phys.\ Rev.\  {\bf D49}, 2729-2733 (1994).
  [hep-ph/9301240].
  
\bibitem{Weinberg:1974hy} 
  S.~Weinberg,
  ``Gauge and Global Symmetries at High Temperature,''
  Phys.\ Rev.\ D {\bf 9}, 3357 (1974).
  doi:10.1103/PhysRevD.9.3357

  R.~N.~Mohapatra and G.~Senjanovi\'c,
  ``Broken Symmetries at High Temperature,''
  Phys.\ Rev.\ D {\bf 20}, 3390 (1979).
  doi:10.1103/PhysRevD.20.3390
  
  R.~N.~Mohapatra and G.~Senjanovi\'c,
  ``High Temperature Behavior of Gauge Theories,''
  Phys.\ Lett.\ B {\bf 89}, 57 (1979).
  doi:10.1016/0370-2693(79)90075-3
  
\bibitem{Dvali:1995cc} 
  G.~R.~Dvali and G.~Senjanovi\'c,
  ``Is there a domain wall problem?,''
  Phys.\ Rev.\ Lett.\  {\bf 74}, 5178 (1995)
  doi:10.1103/PhysRevLett.74.5178
  [hep-ph/9501387].
%
 
  G.~R.~Dvali, A.~Melfo and G.~Senjanovi\'c,
  ``Nonrestoration of spontaneously broken P and CP at high temperature,''
  Phys.\ Rev.\ D {\bf 54}, 7857 (1996)
  doi:10.1103/PhysRevD.54.7857
  [hep-ph/9601376].
  
\bibitem{Chang:1983fu} 
  D.~Chang, R.~N.~Mohapatra and M.~K.~Parida,
  ``Decoupling Parity and SU(2)-R Breaking Scales: A New Approach to Left-Right Symmetric Models,''
  Phys.\ Rev.\ Lett.\  {\bf 52}, 1072 (1984).
  doi:10.1103/PhysRevLett.52.1072
  
   
\bibitem{delAguila:1980qag}
  F.~del Aguila and L.~E.~Ibanez,
  ``Higgs Bosons in SO(10) and Partial Unification,''
  Nucl.\ Phys.\ B {\bf 177}, 60 (1981).
  doi:10.1016/0550-3213(81)90266-2

  T.~G.~Rizzo and G.~Senjanovi\'c,
  ``Grand Unification and Parity Restoration at Low-energies. 2. Unification Constraints,''
  Phys.\ Rev.\ D {\bf 25}, 235 (1982).
  doi:10.1103/PhysRevD.25.235
%
 
   
%
\bibitem{Beall:1981ze} 
  G.~Beall, M.~Bander and A.~Soni,
  ``Constraint on the Mass Scale of a Left-Right Symmetric Electroweak Theory from the K(L) K(S) Mass Difference,''
  Phys.\ Rev.\ Lett.\  {\bf 48}, 848 (1982).
  
 
\bibitem{Ferrari:2000sp}
  A.~Ferrari {\it et al.},
  ``Sensitivity study for new gauge bosons and right-handed Majorana neutrinos
  in p p collisions at s = 14-TeV,''
  Phys.\ Rev.\  D {\bf 62}, 013001 (2000).

\bibitem{Gninenko:2006br} 
  S.~N.~Gninenko, M.~M.~Kirsanov, N.~V.~Krasnikov and V.~A.~Matveev,
  Phys.\ Atom.\ Nucl.\  {\bf 70}, 441 (2007).
  doi:10.1134/S1063778807030039
 
   
\bibitem{Khachatryan:2014dka} 
  V.~Khachatryan {\it et al.} [CMS Collaboration],
  ``Search for heavy neutrinos and $\mathrm {W}$ bosons with right-handed couplings in proton-proton collisions at $\sqrt{s} = 8\,\text {TeV} $,''
  Eur.\ Phys.\ J.\ C {\bf 74}, no. 11, 3149 (2014)
  doi:10.1140/epjc/s10052-014-3149-z
  [arXiv:1407.3683 [hep-ex]].


\bibitem{Tello:2010am}
  V.~Tello, M.~Nemev\v{s}ek, F.~Nesti, G.~Senjanovi\'c and F.~Vissani,
  ``Left-Right Symmetry: from LHC to Neutrinoless Double Beta Decay,''
  Phys.\ Rev.\ Lett.\  {\bf 106} (2011) 151801
  doi:10.1103/PhysRevLett.106.151801
  [arXiv:1011.3522 [hep-ph]].
  
  M.~Nemev\v{s}ek, F.~Nesti, G.~Senjanovi\'c and V.~Tello,
  ``Neutrinoless Double Beta Decay: Low Left-Right Symmetry Scale?,''
  arXiv:1112.3061 [hep-ph].
  
See also,
  P.~S.~Bhupal Dev, S.~Goswami, M.~Mitra and W.~Rodejohann,
  ``Constraining Neutrino Mass from Neutrinoless Double Beta Decay,''
  Phys.\ Rev.\ D {\bf 88}, 091301 (2013)
  doi:10.1103/PhysRevD.88.091301
  [arXiv:1305.0056 [hep-ph]].

  W.~C.~Huang and J.~Lopez-Pavon,
  ``On neutrinoless double beta decay in the minimal left-right symmetric model,''
  Eur.\ Phys.\ J.\ C {\bf 74}, 2853 (2014)
  doi:10.1140/epjc/s10052-014-2853-z
  [arXiv:1310.0265 [hep-ph]].

\bibitem{Casas:2001sr} 
  J.~A.~Casas and A.~Ibarra,
  ``Oscillating neutrinos and $\mu \to$ e  $\gamma$,''
  Nucl.\ Phys.\ B {\bf 618}, 171 (2001)
  [hep-ph/0103065].



\bibitem{Pilaftsis:1991ug}
  A.~Pilaftsis,
  ``Radiatively induced neutrino masses and large Higgs neutrino couplings in the standard model with Majorana fields,''
  Z.\ Phys.\ C {\bf 55} (1992) 275
  [hep-ph/9901206].
  
%


\bibitem{Buchmuller:1991tu} 
  W.~Buchmuller and C.~Greub,
  ``Heavy Majorana neutrinos in electron - positron and electron - proton collisions,''
  Nucl.\ Phys.\ B {\bf 363}, 345 (1991).
  doi:10.1016/0550-3213(91)80024-G
  

\bibitem{Zhang:2007fn} 
  Y.~Zhang, H.~An, X.~Ji and R.~N.~Mohapatra,
  ``Right-handed quark mixings in minimal left-right symmetric model with general CP violation,''
  Phys.\ Rev.\ D {\bf 76}, 091301 (2007)
  doi:10.1103/PhysRevD.76.091301
  [arXiv:0704.1662 [hep-ph]].
 
\bibitem{Gunion:1986im}
  J.~F.~Gunion, B.~Kayser, R.~N.~Mohapatra, N.~G.~Deshpande, J.~Grifols, A.~Mendez, F.~I.~Olness and P.~B.~Pal,
  PRINT-86-1324 (UC,DAVIS).
  
  J.~F.~Gunion, H.~E.~Haber, G.~L.~Kane and S.~Dawson,
  Front.\ Phys.\  {\bf 80}, 1 (2000).
 
  For a recent study on the LHC feasibility, see A.~Maiezza, M.~Nemev\v{s}ek and F.~Nesti,
  ``Lepton Number Violation in Higgs Decay at LHC,''
  Phys.\ Rev.\ Lett.\  {\bf 115}, 081802 (2015)
  doi:10.1103/PhysRevLett.115.081802
  [arXiv:1503.06834 [hep-ph]].
  
 Also, M.~Nemev\v{s}ek, F.~Nesti and J.~C.~Vasquez, to appear.
 
\bibitem{Dev:2016dja} 
  P.~S.~B.~Dev, R.~N.~Mohapatra and Y.~Zhang,
  ``Probing the Higgs Sector of the Minimal Left-Right Symmetric Model at Future Hadron Colliders,''
  JHEP {\bf 1605}, 174 (2016)
  doi:10.1007/JHEP05(2016)174
  [arXiv:1602.05947 [hep-ph]].
 
  
\bibitem{Zhang:2007da} 
  Y.~Zhang, H.~An, X.~Ji and R.~N.~Mohapatra,
  ``General CP Violation in Minimal Left-Right Symmetric Model and Constraints on the Right-Handed Scale,''
  Nucl.\ Phys.\ B {\bf 802}, 247 (2008)
  [arXiv:0712.4218 [hep-ph]].
  
  A.~Maiezza, M.~Nemev\v sek, F.~Nesti, G.~Senjanovi\'c,
  ``Left-Right Symmetry at LHC,''
  Phys.\ Rev.\  {\bf D82 } (2010)  055022.
  [arXiv:1005.5160 [hep-ph]].
  
  S.~Bertolini, A.~Maiezza and F.~Nesti,
  ``Present and Future K and B Meson Mixing Constraints on TeV Scale Left-Right Symmetry,''
  Phys.\ Rev.\ D {\bf 89}, no. 9, 095028 (2014)
  [arXiv:1403.7112 [hep-ph]].

 
\bibitem{Blanke:2011ry} 
  M.~Blanke, A.~J.~Buras, K.~Gemmler and T.~Heidsieck,
  ``$\Delta F = 2$ observables and $B \to X_q$ gamma decays in the Left-Right Model: Higgs particles striking back,''
  JHEP {\bf 1203}, 024 (2012)
  doi:10.1007/JHEP03(2012)024
  [arXiv:1111.5014 [hep-ph]].

  
\bibitem{Maiezza:2016bzp} 
  A.~Maiezza, M.~Nemev\v sek and F.~Nesti,
  ``Perturbativity and mass scales of Left-Right Higgs bosons,''
  arXiv:1603.00360 [hep-ph].
  
  A.~Maiezza, G.~Senjanovi\'c and J.~C.~Vasquez,
 ``Higgs sector of the Left-Right symmetric theory,'' 
 to appear.

%


%
   
\bibitem{Maiezza:2014ala} 
  A.~Maiezza and M.~Nemev\v sek,
  ``Strong P invariance, neutron electric dipole moment, and minimal left-right parity at LHC,''
  Phys.\ Rev.\ D {\bf 90}, no. 9, 095002 (2014)
  [arXiv:1407.3678 [hep-ph]].
 
%


\bibitem{Nemevsek:2011hz} 
  M.~Nemev\v sek, F.~Nesti, G.~Senjanovi\'c and Y.~Zhang,
  ``First Limits on Left-Right Symmetry Scale from LHC Data,''
  Phys.\ Rev.\ D {\bf 83}, 115014 (2011)
  [arXiv:1103.1627 [hep-ph]].


\bibitem{Das:2012ii} 
  S.~P.~Das, F.~F.~Deppisch, O.~Kittel and J.~W.~F.~Valle,
  ``Heavy Neutrinos and Lepton Flavour Violation in Left-Right Symmetric Models at the LHC,''
  Phys.\ Rev.\ D {\bf 86}, 055006 (2012)
  doi:10.1103/PhysRevD.86.055006
  [arXiv:1206.0256 [hep-ph]].
  
\bibitem{Vasquez:2014mxa} 
  J.~C.~Vasquez,
  ``Right-handed lepton mixings at the LHC,''
  arXiv:1411.5824 [hep-ph].



\bibitem{Han:2012vk} 
  T.~Han, I.~Lewis, R.~Ruiz and Z.~g.~Si,
  ``Lepton Number Violation and $W^\prime$ Chiral Couplings at the LHC,''
  Phys.\ Rev.\ D {\bf 87}, 035011 (2013)
  [Erratum-ibid.\ D {\bf 87}, no. 3, 039906 (2013)]
  [arXiv:1211.6447 [hep-ph]].

\bibitem{Gluza:2016qqv} 
  J.~Gluza, T.~Jelinski and R.~Szafron,
  ``Lepton number violation and ÔDiracnessÕ of massive neutrinos composed of Majorana states,''
  Phys.\ Rev.\ D {\bf 93}, no. 11, 113017 (2016)
  doi:10.1103/PhysRevD.93.113017
  [arXiv:1604.01388 [hep-ph]].
 

 

\end{thebibliography}
\end{document}